\documentstyle[11pt,appb2]{article}
\begin{document}
\begin{flushleft}
DESY 98-70 \hfill {\tt hep-ph/9806355}\\
June 1998
\end{flushleft}

\def\refname{\large{\bf References}}
\eqsec
\begin{center}
{\Large\bf
Relations among polarized and unpolarized splitting functions
beyond leading order\footnote{Presented at the conference 
"Loops and Legs in
Gauge Theories" at Rheinsberg, Germany, 19-24  April, 1998. \\
Work supported in part by EU contract FMRX-CT98-0194.}}

\vspace{5mm}
{\sc
J. Bl\"umlein, V. Ravindran and W.L. van Neerven
\footnote {\rm
On leave of absence from Instituut-Lorentz, 
University of Leiden, P.O. Box 9506, 2300 HA Leiden, The Netherlands.}}

\vspace{5mm}
{DESY-Zeuthen, 6 Platenenalle, D-15738 Zeuthen, Germany.}
\end{center}

\nopagebreak

\abstract{\noindent The role of various symmetries in the evaluation of
splitting functions and coefficient functions is discussed.  
The scale invariance in hard processes is known to be a guiding tool
to understand the dynamics. We discuss the constraints on splitting functions 
coming from various symmetries such as scale, 
conformal and supersymmetry.  We also discuss the Drell-Levy-Yan relation 
among splitting and coefficient functions in various schemes. The
relations coming from conformal symmetry are also presented.}
  
\section{Scale Transformation}
\noindent Symmetries are known to be very useful guiding tool to understand the 
dynamics of various
physical phenomena.  Particularly, continuous symmetries played
an important role in particle physics to unravel the structure of dynamics
at low as well as high energies.  In hadronic physics, such symmetries
at low energies were found to be useful to classify various hadrons.  
At high energy, where the masses of the particles can be neglected, one
finds in addition to the above mentioned symmetries
new symmetries such as conformal and scale invariance. This for instance
happens in deep inelastic lepton-hadron scattering (DIS) where
the energy scale is much larger than the hadronic mass scale.
At these energies one can in principle ignore the mass scale and the resulting
dynamics is purely scale independent. 

Limiting ourselves to scale transformations the latter is defined by\\
$x_\mu \rightarrow e^t~ x_\mu$.  An arbitrary quantum field $\hat \phi (x)$
is then transformed as follows
\begin{eqnarray}
\label{eq1}
\hat \phi (x) \rightarrow U^\dagger~ \hat \phi(x)~ U=
e^{d_0 t}~ \hat \phi(e^{-t} x) \,,
\end{eqnarray}
where $U$ is the unitary operator and $d_0$ is its 
canonical dimension. Under this transformation the $n$-point Green's function
$E_n(p_i,g)$ behaves like
\begin{eqnarray}
\label{eq2}
E_n(e^t p,g) =E_n(p,g) e^{t(D-n~d_0)} \,,
\end{eqnarray}
where $p_i$ are the momenta and $g$ denotes the coupling constant. 
However in perturbation theory, like QCD, scale invariance is broken due
to the introduction of a regulator scale which is rigid under conformal
and scale transformation. Even if the regulator is removed in the renormalized
Green's funstion a renormaliztion scale $\mu$ is left which is rigid too.
In this case the Green's function does not satisfy a simple scaling 
equation anymore and the latter is replaced by the
Callan Symmanzik (CS) equation \cite{callan} which
reads
\begin{eqnarray} 
\label{eq3}
\left[\frac{\partial}{\partial t}+\beta(g)\frac{\partial}{\partial g} -D 
   +n~(d_0+\gamma(g))  \right] E_n(e^t p_i,g,\mu)=0 \,,
\end{eqnarray}
where $\beta(g)$ and $\gamma(g)$ denote the beta-function and the anomalous
dimension respectively with the property $\beta \rightarrow 0$ and 
$\gamma \rightarrow 0$ as $g \rightarrow 0$. 
In the case $\beta(g_c)=0$ at some
fixed point $g=g_c\not =0$  scale invariance is restored and
the solution to this equation becomes
\begin{eqnarray}
\label{eq4}
E_n(e^t p,g_c,\mu)=E_n(p,g_c,\mu) e^{t(D-n(d_0+\gamma(g_c))} \,.
\end{eqnarray}
Let us discuss the beta-function and the anomalous dimensions of composite
operators for QCD. The latter are derived from the Green's function
\begin{eqnarray}
\label{eq5}
&& E^{(n)}_{ij}(p,g,\mu)= 
\nonumber\\[2ex]
&& \int d^4x_1\,d^4x_2\, e^{ip.(x_1-x_2)} \langle 0 \mid
T\Big ( \hat \phi_j (x_1) O_i^{(n)}(0) \hat \phi_j (x_2)\Big ) \mid 0 
\rangle \,,
\end{eqnarray}
Here $ O_i^{(n)}$ denotes the composite operator of spin $n$ which is build
out of quark and gluon fields $\hat \phi_j$ with $i,j=q,g$. If one
chooses $D$-dimensional regularization the renormalized Green's
functions and the bare Green's functions (indicated by the subscript $u$) 
are related by
\begin{eqnarray}
\label{eq6}
E^{(n)}_{ij}(p,g,\mu)=\Big (Z^{(n)}\Big )^{-1}_{il}(\epsilon,g,\mu) 
E^{(n)}_{lj,u}(p,g,\epsilon) \,,
\end{eqnarray}
where $Z_{ij}(\epsilon,g,\mu)$ is the operator renormalization constant and
$\epsilon$ indicates the ultraviolet pole terms in $D$-dimensional
regularization ($D=4-2 \epsilon$). Notice that there is more than one
operator involved in the renormalization so that we have to deal with
mixing. If we amputate the external legs of the Green function in (\ref{eq5})
the anomalous dimension of the composite operator $O^{(n)}$ is given by
\begin{eqnarray}
\label{eq7}
\gamma_{ij}^{(n)}(g)=\beta(g,\epsilon)\Big ( Z^{(n)} \Big )^{-1}_{il}
(g,\epsilon)\frac {d~Z^{(n)}_{lj}(g,\epsilon)}{dg} \,.
\end{eqnarray}
The renormalization constant $Z(\epsilon,g)$ has an expansion in $1/\epsilon$ as
\begin{eqnarray}
\label{eq8}
Z^{(n)}_{ij}(\epsilon,g)=1+\frac{1}{\epsilon} Z^{(n),(1)}_{ij}(g)
+\frac{1}{\epsilon^2} Z^{(n),(2)}_{ij}(g)+\cdots \,.
\end{eqnarray}
Since the beta-function has the following form
\begin{eqnarray}
\label{eq9}
 \beta(g,\epsilon)=\epsilon \frac{g}{2}+\beta_0 \frac{g^3}{16\pi^2} 
+\cdots \,,
\end{eqnarray}
the $\gamma^{(n)}_{ij}$ are finite in the limit $\epsilon \rightarrow 0$ 
and one gets
\begin{eqnarray}
\label{eq10}
\gamma{(n)}_{ij}(g)=\frac{g}{2} \Big (Z^{(n),(1)}\Big )^{-1}_{il}
\frac{d~Z^{(n),(1)}_{lj}}{dg} \,.
\end{eqnarray}
For the amputated Green's function the CS equation (\ref{eq3}) reads
\begin{eqnarray}
\label{eq11}
\left[\mu \frac{\partial}{\partial \mu}+\beta(g)\frac{\partial}{\partial g} 
+\gamma^{(n)}_{il}(g)\right]
E_{lj}^{(n)}(p,g,\mu)=0 \,.
\end{eqnarray}
In the case of scale invariance i.e. $\beta(g_c)=0$ and no mixing the above
CS equation has the simple solution
\begin{eqnarray}
\label{eq12}
E^{(n)}(\mu^2)=E^n(\mu_0^2)\left(\frac{\mu^2}{\mu_0^2}\right)^{\gamma^{(n)}} 
\,,
\end{eqnarray}
with
\begin{eqnarray}
\label{eq13}
\gamma^{(n)}(\alpha_s)=\left(\frac{\alpha_s}{2 \pi}\right) \gamma^{(n),(0)}
+\left(\frac{\alpha_s}{2 \pi}\right)^2 \gamma^{(n),(1)} +\cdots 
\qquad \alpha_s \equiv \frac{g^2}{4\pi} \,.
\end{eqnarray}
The splitting functions $P(x,\alpha_s)$ are related to these anomalous 
dimensions via a Mellin transformation given by
\begin{eqnarray}
\label{eq14}
\gamma^{(n),(i)}(\alpha_s)=\int_0^1 dx\, x^{n-1} P^{(i)}(x,\alpha_s) 
\qquad i=0,1,2 \cdots \,.
\end{eqnarray}
The above analysis based on scale transformation suggests that only in a scale 
invariant theory,
the Green's function has the form given in the Eq. (\ref{eq12}). This 
will be no longer true in a scale breaking theory like QCD. The same
will hold for the anomalous dimension which in the case of no mixing and
scale invariance is independent of the subtraction scheme. This will change
when this symmetry is broken as we will show below.

\section{Supersymmetric Relations}
\noindent In this section we discuss some relations among splitting functions 
which govern the evolution of quark and gluon parton densities. These
relations are valid when QCD becomes a supersymmetric 
${\cal N}=1$ gauge field theory where both quarks and gluons are put
in the adjoint representation with respect to the local gauge symmetry
$SU(N)$. In this case one gets a simple relation between the colour factors
which become $C_F=C_A=2 T_f = N$.
In the case of spacelike splitting functions, which govern
the evolution of the parton densities in deep inelastic lepton-hadron
scattering, one has made the claim (see \cite{dokshitser}) that the
combination defined by 
\begin{eqnarray}
\label{eq15}
{\cal R}^{(i)}= P_{qq}^{(i)}- P_{gg}^{(i)}+ P_{gq}^{(i)}- P_{qg}^{(i)} \,,
\end{eqnarray}
is equal to zero, i.e., ${\cal R}^{(i)}=0$. This relation should follow
from an ${\cal N}=1$ supersymmetry although no explicit proof has been given
yet. An explicit calculation at leading order(LO) confirms this claim so
that we have ${\cal R}^{(0)}=0$
However at next to leading Order(NLO), when these splitting functions are 
computed in the ${\overline {\rm MS}}$-scheme, it turns out that 
${\cal R}_{\overline {\rm MS}}^{(1)}\not =0$. Actually one finds in the unpolarized
case (see \cite{curci})
\begin{eqnarray}
\label{eq16}
{\cal R}_{\overline {\rm MS}}^{\rm unpol,(1)}&=&-\frac{5}{6}
-\frac{2}{3 x}+\frac{23 x}{3}-7 x^2-4 \delta(1-x)
\nonumber\\[2ex]
&&-(1-2x-4x^2)\ln(x) \,,
\end{eqnarray}
whereas in the polarized case \cite{neerven} one obtains
\begin{eqnarray}
\label{eq17}
{\cal R}_{\overline {\rm MS}}^{\rm pol,(1)}=\frac{28}{3}
-\frac{44x}{ 3}-8 (1-x)\ln(x)-4 \delta(1-x) \,.
\end{eqnarray}
The reason that this relation is violated can be attributed to the 
regularization method and the 
renormalization scheme in which these splitting functions are computed.
In this case it is $D$-dimensional regularization and the $\overline{\rm MS}$-scheme 
which breaks the supersymmetry.  In fact, the breaking occurs already at the
$\epsilon$ dependent part of the leading order splitting functions. Although
this does not affect the leading order splitting functions in the limit 
$\epsilon \rightarrow 0$ it leads to
a finite contribution at the NLO level via the $1/\epsilon^2$ terms which
are characteristic of a two-loop calculation (see Eq. (\ref{eq8})).
If one carefully removes such breaking terms at the LO level consistently,
one can avoid these terms at NLO level. They can be
avoided if one uses $D$-dimensional reduction which preserves the supersymmetry. 
An other possibility is that one can convert the splitting functions from one
scheme to another by the following transformation
\begin{eqnarray}
\label{eq18}
Z^{(n)} \rightarrow  Z^{(n)'}=\hat Z_F Z^{(n)} \,,
\end{eqnarray}
where $\hat Z_F$ is a finite renormalization.
Under this transformation the anomalous dimensions in the new scheme become
\begin{eqnarray}
\label{eq19}
\gamma^{(n)} \rightarrow  \gamma^{(n)'}
=\hat Z_F^{-1} \gamma^{(n)} \hat Z_F-\beta(g)
\displaystyle{\frac{d\hat Z_F^{-1}}{dg}}\hat Z_F \,.
\end{eqnarray}
After Mellin inversion see (\ref{eq14}) one gets in the unpolarized case 
\begin{eqnarray}
\label{eq20}
\hat Z_F^{\rm unpol}=
\left(
\begin{array}{cc}
-2+2 x +\delta (1-x)&0\\
-2 x& -4 x +4 x^2 +\frac{1}{3} \delta(1-x)
\end{array}
\right) \,,
\end{eqnarray}
and for the polarized case we have
\begin{eqnarray}
\label{eq21}
\hat Z_F^{\rm pol}=
\left(
\begin{array}{cc}
-2+2 x +\delta(1-x)&0\\
0&\frac{1}{3} \delta(1-x)
\end{array}
\right) \,.
\end{eqnarray}
In this new (primed) scheme it turns out that ${\cal R}^{(1)'}=0$ 

The above observations also apply to the timelike splitting functions, 
denoted by a tilde, which govern the evolution of fragmentation functions.
Substitution of their expressions \cite{furmanski} into Eq. (\ref{eq15}) 
yields the $\overline{\rm MS}$-scheme results
\begin{eqnarray}
\label{eq22}
\tilde {\cal R}^{\rm unpol,(1)}=\frac{5 x}{3}-\frac{2}{3 x}+\frac{13}{6}
- x^2-\frac{1}{2} \delta(1-x) -(1-2x-4x^2)\ln(x) \,.
\end{eqnarray}
For the polarized case we need the splitting functions in \cite{stratmann}
so that we get
\begin{eqnarray}
\label{eq23}
\tilde {\cal R}^{\rm pol,(1)}=\frac{11}{6}-\frac{7x}{6}  
+ (1-x)\ln(x)-\frac{1}{2} \delta(1-x) \,.
\end{eqnarray}
It appears that the scheme transformation, introduced for the 
spacelike case in Eqs. (\ref{eq20})
(\ref{eq21}), or the use of supersymmetric reduction also lead in the timelike 
case to the result $\tilde {\cal R}^{(1)'}=0$ in Eq. (\ref{eq15}). 

\section{Drell-Levy-Yan Relation}
\noindent The Drell-Levy-Yan relation (DLY) \cite{drell} relates the  structure  functions \\
$F(x,Q^2)$ measured in deep inelastic scattering to the fragmentation functions 
$\tilde F(\tilde x,Q^2)$ observed 
in $e^+~e^-$-annihilation. Here $x$ denotes the Bj{\o}rken
scaling variable which in deep inelastic scattering and $e^+~e^-$-annihilation
is defined by $x=Q^2/2p.q$ and $\tilde x=2p.q/Q^2$ respectively.
Notice that in deep inelastic scattering the virtual photon
momentum $q$ is spacelike i.e. $q^2=-Q^2<0$ whereas in $e^+~e^-$-annihilation it
becomes timelike $q^2=Q^2>0$. Further $p$ denotes the in or outgoing hadron
momentum.
The DLY relation looks as follows 
\begin{eqnarray}  
\label{eq24}
\tilde F_i(\tilde x,Q^2)= x {\cal A }c\left[F_i(1/x,Q^2)\right] \,,
\end{eqnarray}
where ${\cal A }c$ denotes the analytic continuation from the region
$0<x \le 1$ (DIS) to $1<x<\infty$ (annihilation region). 
At the level of splitting functions we have
\begin{eqnarray}  
\label{eq25}
\tilde P_{ij}(\tilde x)= x {\cal A }c\left[P_{ji}(1/x)\right] \,.
\end{eqnarray}
At LO, one finds  $\tilde P^{(0)}(\tilde x)=P^{(0)T}(x)$. Explicity,
\begin{eqnarray}  
\label{eq26}
\tilde P_{qq}^{(0)}(\tilde x)= -x P_{qq}^{(0)}(1/x) \qquad
\tilde P_{qg}^{(0)}(\tilde x)=\frac{2 T_f}{C_F} x P_{gq}^{(0)}(1/x) \,,
\nonumber\\[2ex] 
\tilde P_{gq}^{(0)}(\tilde x)= \frac{C_F}{2 T_f}x P_{qg}^{(0)}(1/x) \qquad
\tilde P_{gg}^{(0)}(\tilde x)= -x P_{gg}^{(0)}(1/x) \,.
\end{eqnarray}
At the leading order level, one finds that 
\begin{equation}
 \tilde P_{qq}^{(0)}(\tilde x) =
 P_{qq}^{(0)}(x)
\end{equation}
, which is nothing but Gribov-Lipatov relation \cite {gribov}.  This relation
in terms of physical observables is known to be violated when one goes beyond leading order \cite{curci}.  
On the other hand the DLY (analytical continuation) relation defined above holds at the level of
physical quatities provided the analytical continuation is performed in both $x$ as well as
the scale $Q^2$ ($Q^2 \rightarrow -Q^2$) (see below).

In analytical continuation, care is needed when one goes beyond LO when dimensional regularization is adopted. 
The correct 
${\cal A }c$ relation in DR scheme reads as follows \cite{curci}:
\begin{eqnarray}  
\label{eq27}
\tilde P_{ij}(\tilde x)= x^{1-2 \epsilon} {\cal A }c\left[P_{ji}(1/x)\right] 
\,. 
\end{eqnarray}
The extra term $x^{-2 \epsilon}$ arises due to the difference between the 
spacelike and timelike phase space integrations. Starting
from the definitions of splitting functions,
\begin{eqnarray}
\label{eq28}
P(x)=\beta(\alpha_s,\epsilon) \frac{d \ln Z(x,\alpha_s,\epsilon)}
{d \ln \alpha_s} \qquad
\tilde P(\tilde x)=\beta(\alpha_s,\epsilon) 
\frac{d \ln \tilde Z(\tilde x,\alpha_s,\epsilon)} {d \ln \alpha_s} \,,
\end{eqnarray}
and
\begin{eqnarray}
\label{eq29}
x^{-2 \epsilon} {\cal A }c\left[Z(1/x,\epsilon)\right]=
Z_F(x) {\cal A }c\left[Z(1/x,\epsilon)\right] \,,
\end{eqnarray}
one finds that the splitting functions are related by simple relation
\begin{eqnarray}
\label{eq30}
\tilde P(\tilde x)=x {\cal A }c\left[P(1/x)\right]+
  \mbox{contributions coming from} \, Z_F \,.
\end{eqnarray}
The DLY relation between NLO coefficient functions appearing in
DIS and $e^+e^-$ can be worked out in the same way as we did for the
splitting functions above. In the subsequent part of this section we will 
only study the gluonic
coefficient functions corresponding to the deep inelastic structure
functions and the fragmentation functions.
The conclusions also apply to the quark coefficient functions as well.

\noindent The spacelike gluonic coefficient function for the polarized case in DIS
originates from
photon-gluon fusion process and is given by \cite{disg}  
\begin{eqnarray}
\label{eq31}
{\cal C}_{1,g}(x,Q^2)|_{\overline {\rm MS}}
=e_q^2 \frac{\alpha_s}{4 \pi}\left((2 x -1) 
\ln\left(\frac{Q^2 (1-x)}{\mu^2 x}\right) +3- 4 x\right) \,.
\end{eqnarray}
In the above, the collinear singularity is treated in $D$-dimensional 
regularization and
the scale $\mu$ is the factorization scale. For $e^+~e^-$-annihilation 
the timelike coefficient function becomes \cite{eeg}
\begin{eqnarray}
\label{eq32}
\tilde {\cal C}_{1,g}(\tilde x,Q^2)|_{\overline {\rm MS}}&
=&x {\cal C}_{1,g}(1/x,Q^2)|_{\overline {\rm MS}}
+ 2\,P_{gq}^{(0)} \ln(x) \,.
\end{eqnarray}
\noindent The violation of DLY relation is due to the regularization method
and the scheme we have adopted to remove the collinear singularities 
from the partonic cross sections. 
This is the reason we get a mismatch between the phase space integrations
in the spacelike and timelike case which is equal to $x^{-2 \epsilon}$.
This factor is multiplied with the lowest order pole term which leads
to the finite contribution on the right hand side of Eq. (\ref{eq32}).

\noindent The violation is an artifact of dimensional regularization and the
choice of the $\overline {\rm MS}$-scheme.  For example if one chooses
a regularization where the gluon gets a mass $m_g$ and one removes
the mass singularity $\ln(\mu^2/m_g^2)$ only, the space-and timelike coefficient
functions become \cite{ravi}
\begin{eqnarray}
\label{eq33}
{\cal C}_{1,g}(x,Q^2)|_{m_g\not=0} =
e_q^2 \frac{ \alpha_s}{4 \pi}(2 x -1)\left( 
\ln\left(\frac{Q^2}{\mu^2 x^2}\right) -2\right) \,,
\end{eqnarray}
and
\begin{eqnarray}
\label{eq34}
\tilde {\cal C}_{1,g}(\tilde x,Q^2)|_{m_g\not=0}=  
x {\cal C}_{1,g}(1/x,Q^2)|_{m_g\not=0} \,,
\end{eqnarray}
respectively so that the DLY relation is satisfied. The same happens
when the quark gets a mass $m_q$. After removing the mass singularity
$\ln(\mu^2/m_q^2)$ one gets \cite{bluemlein}
\begin{eqnarray}
\label{eq35}
{\cal C}_{1,g}(x,Q^2)|_{m_q\not=0}
= e_q^2 \frac{\alpha_s}{4 \pi}\left((2 x -1) 
\ln\left(\frac{Q^2 (1-x)}{\mu^2 x}\right) +3- 4 x\right) \,,
\end{eqnarray}
and
\begin{eqnarray}
\label{eq36}
\tilde {\cal C}_{1,g}(\tilde x,Q^2)|_{m_q\not=0}
=  x {\cal C}_{1,g}(1/x,Q^2)|_{m_q\not=0} \,.
\end{eqnarray}
Hence the violation of the DLY relation for the splitting functions and
the coefficient functions separately is just
an artifact of the adopted regularization method and the subtraction scheme.
When these coefficient functions are combined with the splitting functions
in a scheme invariant way as for instance happens for the structure
functions  and fragmentation functions the above relation holds.
The reason for the cancellation of the DLY violating terms among 
the splitting functions and coefficient functions is that the former
are generated by simple scheme transformations. 

\section {Supersymmetric and Conformal Relations}
\noindent In this section we study the constraints coming from the conformal symmetry
on the splitting functions in an ${\cal N}=1$ supersymmetry. The following
set of relations have been derived \cite{bukhvostov} between the unpolarized
($P_{ij}$) and polarized ($\Delta P_{ij}$) splitting functions.
\begin{eqnarray}
\label{eq37}
\left (P_{qq}-P_{qg} \right) +\left(\Delta P_{qq}-\Delta P_{qg}\right)&=&
x \left(P_{qq}+P_{gq}+\Delta P_{qq}+\Delta P_{gq} \right) 
\end{eqnarray}
\begin{eqnarray}
\label{eq38}
\left (P_{qq}-P_{qg} \right) -\left(\Delta P_{qq}-\Delta P_{qg}\right)&=&
-x \left(P_{qq}+P_{gq}-\Delta P_{qq}-\Delta P_{gq} \right)
\end{eqnarray}
The LO splitting functions satisfy the above relations but at NLO level they are
violated in the $\overline {\rm MS}$-scheme. In the latter scheme the difference
between the left- and righthand side of Eqs. (\ref{eq37}) and (\ref{eq38}) 
is given by
\begin{eqnarray}
\label{eq39}
\frac{1}{3} - 8 x +\frac{29 x^2}{6}+\frac{4 x^3}{3}-2 x \ln(x)-5 x^2 \ln(x) \,,
\end{eqnarray}
and
\begin{eqnarray}
\label{eq40}
-2 +\frac{8x}{3} +\frac{13 x^2}{6}-\frac{4 x^3}{3}-2 \ln(x)
- 2 x \ln(x)- x^2 \ln(x) \,,
\end{eqnarray}
respectively. Following the discussion below Eq. (\ref{eq15}) 
these relations can be preserved by making finite scheme transformations. 
Another interesting relation in \cite{bukhvostov} is the one between 
the non-diagonal entries of splitting
function matrix:
\begin{eqnarray}
\label{eq41}
x \frac{d}{dx}\left((\Delta) P_{qg}-(\Delta) P_{gq}\right)
= 2 (\Delta) P_{qg}+(\Delta) P_{gq}
\end{eqnarray}
The known LO splitting functions satisfy this relation but it is violated by
NLO splitting functions in $\overline {\rm MS}$ scheme. Interestingly, 
the violation comes from the terms
such as $\ln(x) \ln(1-x)$.  These terms can not be removed by finite scheme
transformation so that the above equation does not hold anymore in NLO
irrespective of the chosen scheme.
\section{Conclusions}
\noindent We have discussed the relations between the splitting functions
coming from various symmetries such as
scale symmetry, conformal symmetry and supersymmetry on NLO splitting functions
and coefficient functions.  The Drell-Levy-Yan relation among them
is also discussed at NLO level.  Most of the relations coming from these 
symmetries
are violated in dimensional regularization with $\overline {\rm MS}$ prescription.
The breaking terms can be identified at the leading order level, and by simple
finite renormalization, one can preserve the relations coming from scale and
supersymmetric constraints. The breaking due to conformal non-invariant
terms (see Eq. (\ref{eq41})) can not be cured by a simple finite 
renormalization.

\end{document}